\documentclass[12pt]{article}
\usepackage{amssymb,amsmath,epsfig}
\begin{document}
\title{\bf Locally Rotationally
Symmetric Bianchi Type $I$ Cosmology in $f(R,T)$ Gravity}

\author{M. Farasat Shamir\thanks{farasat.shamir@nu.edu.pk}\\\\
Department of Sciences and Humanities, \\National University of
Computer and Emerging Sciences,\\ Lahore Campus, Pakistan.}

\date{}

\maketitle
\begin{abstract}
This manuscript is devoted to investigate Bianchi Type $I$
universe in the context of $f(R,T)$ gravity. For this purpose,
we explore the exact solutions of locally rotationally
symmetric Bianchi type $I$ spacetime. The modified field equations are solved by assuming expansion scalar
$\theta$ proportional to shear scalar $\sigma$ which gives
$A=B^n$, where $A,~B$ are the metric coefficients and $n$ is an arbitrary constant.
In particular, three solutions have been found and physical quantities are
calculated in each case.
\end{abstract}


\section{Introduction}

Recent observations from astrophysical data have unfolded an amazing 
picture of expanding universe. The cosmic acceleration is well
supported by high red-shift supernovae, cosmic microwave background anisotropy and galaxy clustering \cite{1}.
Universe seems to be filled with exotic cosmic fluid known as dark energy having strong negative pressure.
It constitutes almost $70$ $\%$ of the total energy budget of our universe.
We can describe dark energy with an equation of state (EoS) parameter $\omega=p/\rho$,
where $\rho$ and $p$ represent the energy density and pressure of dark energy.
It has been proved that the expansion of the universe is accelerating when $w\approx-1$ \cite{nature}.
The phantom like dark energy is found to be in the region where $\omega<-1$. The universe with phantom dark energy
ends up with a finite time future singularity known as cosmic doomsday or big rip \cite{Caldwell}.
The modified theories of gravity seem attractive to explain the phenomenon of
dark energy and late time acceleration. It is now expected that the issues of cosmic acceleration and quintessence 
can be addressed using higher order theories of gravity \cite{Capozziello10}.

Many generalizations of Einstein field equations have been proposed in the last few decades.
$f(T)$ theory of gravity is an example which has been
recently developed. This theory is a generalized version of
teleparallel gravity in which Weitzenb\"{o}ck connection is used
instead of Levi-Civita connection. The theory seems interesting as it may explain the current cosmic acceleration without
involving the dark energy. Some researchers have done a considerable amount of work
in this theory so far \cite{ft}. Another extended theory
known as $f(R)$ theory of gravity has also attracted attention of the researchers in recent years.
The $f(R)$ theory is actually an extension of standard Einstein Hilbert action involving a function of the Ricci
scalar $R$. The cosmic acceleration may be justified by involving the term
$1/R$ which is required at small curvatures. The $f(R)$ theory seems to be
most appropriate due to important $f(R)$ models in cosmological contexts. 

Some viable $f(R)$ gravity models \cite{4} have
been suggested which justify the unification of early-time inflation
and late-time acceleration. The dark matter problems can also be
addressed using viable $f(R)$ gravity models \cite{5}. 
Starobinsky \cite{Starobinsky10} gave first complete inflationary $R+R^2$ model 
viable with the observational data. The cosmological model 
without additional singularities can be constructed using $f(R)$ gravity in 
which both inflation in the early universe and dark energy in the present universe is described \cite{Starobinsky11}.
Hendi and Momeni \cite{bhs} investigated
black hole solutions in $f(R)$ theory of gravity with conformal anomaly.
Jamil et al. to explored $f(R)$ tachyon model using Noether symmetries \cite{Taychon}.
Capozziello et al. \cite{02} have proved that dust matter and dark energy phases can
be achieved by finding the exact solutions using a power law $f(R)$
cosmological model. A reasonable amount of work has been
done so far in this theory \cite{Work}. Some review articles \cite{rev} may
be useful to have a better understanding of the theory.

Recently, Harko et al. \cite{fRT1} proposed a new
generalizations known as $f(R,T)$ theory of gravity in which $R$ is the scalar curvature and $T$ denotes
the trace of the energy-momentum tensor. Jamil et al. \cite{fRT2} reconstructed some cosmological models in $f(R,T)$ gravity
where it was proved that the dust fluid reproduced $\Lambda$CDM.
Sharif and Zubair \cite{fRT529} gave the reconstruction and stability conditions of $f(R,T)$ gravity with Ricci and modified Ricci dark energy.
The same authors \cite{fRT5} discussed the laws of thermodynamics in this
theory. However, it has been established that the first law of black bole thermodynamics is violated for $f(R,T)$ gravity \cite{voilation}.
Santos \cite{godal} investigated G\"{o}del type universe in the context of $f(R,T)$ modified theories of gravity.
Houndjo \cite{fRT4} reconstructed $f(R,T)$ gravity by taking
$f(R,T)=f_1(R)+f_2(T)$ and it was shown that $f(R,T)$ gravity
allowed transition of matter from dominated phase to an acceleration
phase. Harko and Lake \cite{HARKO} investigated
cylindrically symmetric interior string like solutions in $f(R,L_m)$ theory of gravity.
In a recent paper \cite{me144}, we explored the exact solutions of
cylindrically symmetric spacetime in $f(R,T)$ gravity
and recovered two solutions which corresponded to an exterior metric
of cosmic string and a non-null electromagnetic field.

Isotropization is an important issue to discuss whether the universe can result in isotropic solutions without the
need of fine tuning the model parameters \cite{Saridakis}. The universe seems to have an isotropic
and homogeneous geometry at the end of the inflationary
era \cite{Linde}. However, the class of anisotropic geometries has
gained popularity under the light of the recently announced Planck Probe
results \cite{Plank}. Further, it is believed
that the early universe may not have been exactly uniform.
Therefore, inhomogeneous and anisotropic models of universe
have an important role in theoretical cosmology. This
prediction motivates us to describe the early stages of the
universe with the models having anisotropic background. Thus,
the existence of anisotropy in early phases of the universe is an interesting
phenomenon to investigate. Bianchi type models are among the simplest models
with anisotropic background. Many authors \cite{14}
investigated Bianchi type spacetimes in different contexts. Kumar and
Singh \cite{019} solved the field equations in the
presence of perfect fluid using Bianchi type $I$ spacetime in general relativity (GR).
Moussiaux et al. \cite{20} investigated the exact solution for
vacuum Bianchi type $III$ model in the presence of cosmological constant.
Bianchi type $III$ string cosmology with bulk
viscosity has been studied by Xing-Xiang \cite{21}. He assumed expansion scalar proportional to
the shear scalar to find the solutions. Wang \cite{022}
explored string cosmological models in
Kantowski-Sachs spacetime. Magnetized Bianchi type $III$
massive string cosmological models in GR have been investigated by Upadhaya \cite{22}.
Hellaby \cite{023} gave a review of some
recent developments in inhomogeneous models and it was concluded
that the universe is inhomogeneous on many scales.

The investigation of Bianchi type models in modified or alternative
theories of gravity is also another interesting topic of discussion.
Perfect fluid solutions
using Bianchi type $I$ spacetime in scalar tensor theory have been explored by Kumar and Singh \cite{23}. Singh et
al. \cite{24} studied Bianchi type $III$ cosmological models
in scalar tensor theory. Adhav et al. \cite{25} found an exact
solution of the vacuum Brans-Dicke field equations for a spatially homogeneous and anisotropic model.
FRW cosmologies in $f(R)$ gravity have been investigated by Paul et
al. \cite{26}. Bianchi type $I$ model in $f(R)$ gravity was studied where it was shown
how to integrate anisotropic degrees of freedom explicitly and
to reduce the problem to one differential equation for the volume factor \cite{Starobinsky12}.
Vacuum and non-vacuum solutions of Bianchi
types $I$ and $V$ spacetimes in metric $f(R)$ gravity
have been explored \cite{27,28}.
Sharif and Kausar \cite{riz} investigated non-vacuum solutions of Bianchi type $VI$ universe
by considering the isotropic and anisotropic fluids as the source of dark matter and energy.
In a recent paper, we have explored Bianchi type $I$ cosmology in $f(R,T)$ gravity
with some interesting results \cite{me143}. It was concluded that equation of state
parameter $w\rightarrow -1$ as $t\rightarrow\infty$ which suggested an
accelerated expansion of the universe. Thus it is hoped that $f(R,T)$ gravity may explain the resent
phase of cosmic acceleration of our universe. This theory can be
used to explore many issues and may provide some satisfactory
results.

In this paper, we explore the exact solutions
of locally rotationally
symmetric (LRS) Bianchi type $I$ spacetime
in $f(R,T)$ gravity. The field equations are solved by assuming
expansion scalar $\theta$ proportional to shear scalar $\sigma$
which gives $A=B^n$, where $A,~B$ are the metric coefficients and $n$ is
an arbitrary constant. The plan is planned as follows:
Field equations in $f(R,T)$ gravity are briefly
introduced in section \textbf{2}. In section \textbf{3},
the solutions of the field equations are
investigated along with some important physical parameters.
Last section is used to summarize and conclude the results.

\section{$f(R,T)$ Gravity Formalism}

The action for $f(R,T)$ theory of gravity is given by \cite{fRT1}
\begin{equation}\label{1}
S=\int\sqrt{-g}(\frac{1}{2\kappa}f(R,T)+L_{m})d^4x,
\end{equation}
where $g$ is the determinant of the metric tensor $g_{\mu\nu}$ and $L_{m}$ is
the usual matter Lagrangian. It would be worthwhile to mention
that if we replace $f(R,T)$ with $f(R)$, we get the action for
$f(R)$ gravity and replacement of $f(R,T)$ with $R$ leads to the
action of GR.
The $f(R,T)$ gravity field equations are obtained by varying the
action $S$ in Eq.(\ref{1}) with respect to the metric tensor
$g_{\mu\nu}$
\begin{equation}\label{4}
f_R(R,T)R_{\mu\nu}-\frac{1}{2}f(R,T)g_{\mu\nu}-(\nabla_{\mu}
\nabla_{\nu}-g_{\mu\nu}\Box)f_R(R,T)=\kappa
T_{\mu\nu}-f_T(R,T)(T_{\mu\nu}+\Theta_{\mu\nu}),
\end{equation}
where $\nabla_{\mu}$ denotes the covariant derivative and
\begin{equation*}
\Box\equiv\nabla^{\mu}\nabla_{\mu},~~ f_R(R,T)=\frac{\partial
f_R(R,T)}{\partial R},~~ f_T(R,T)=\frac{\partial
f_R(R,T)}{\partial
T},~~\Theta_{\mu\nu}=g^{\alpha\beta}\frac{\delta
T_{\alpha\beta}}{\delta g^{\mu\nu}}.
\end{equation*}
Contraction of Eq.(\ref{4}) yields
\begin{equation}\label{5}
f_R(R,T)R+3\Box f_R(R,T)-2f(R,T)=\kappa T-f_T(R,T)(T+\Theta),
\end{equation}
where $\Theta={\Theta_\mu}^\mu$. This is an important equation
because it provides a relationship between Ricci scalar $R$ and
the trace $T$ of energy momentum tensor.
Using matter Lagrangian $L_m$, the standard matter energy-momentum tensor is derived as
\begin{equation}\label{6}
T_{\mu\nu}=(\rho + p)u_\mu u_\nu-pg_{\mu\nu},
\end{equation}
satisfying the EoS
\begin{equation}\label{eos}
p=w\rho,
\end{equation}
where $u_\mu=\sqrt{g_{00}}(1,0,0,0)$ is the four-velocity in
co-moving coordinates and $\rho$ and $p$ denote energy density
and pressure of the fluid respectively. Perfect fluids problems
involving energy density and pressure are not any easy task to
deal with. Moreover, there does not exist any unique definition
for matter Lagrangian. Thus we can assume the matter Lagrangian as
$L_m=-p$ which gives
\begin{equation}\label{7}
\Theta_{\mu\nu}=-pg_{\mu\nu}-2T_{\mu\nu},
\end{equation}
and consequently the field equations (\ref{4}) take the form
\begin{equation}\label{238}
f_R(R,T)R_{\mu\nu}-\frac{1}{2}f(R,T)g_{\mu\nu}-(\nabla_{\mu}
\nabla_{\nu}-g_{\mu\nu}\Box)f_R(R,T)=\kappa
T_{\mu\nu}+f_T(R,T)(T_{\mu\nu}+pg_{\mu\nu}),
\end{equation}
It is mentioned here that these field equations depend on the
physical nature of matter field. Many theoretical models
corresponding to different matter contributions for $f(R,T)$
gravity are possible. However, Harko et al. \cite{fRT1} gave three classes of
these models
\[ f(R,T)= \left\lbrace
  \begin{array}{c l}
    {R+2f(T),}\\
    {f_1(R)+f_2(T),}\\{f_1(R)+f_2(R)f_3(T).}
  \end{array}
\right. \]\\
In this paper, we consider the first and second class only to explore the exact LRS Bianchi $I$ solutions.

\section{Exact LRS Bianchi Type $I$ Solutions}

Here we first develop some important cosmological parameters
and field equations for LRS Bianchi type $I$ spacetime
and then find the exact solutions of field equations for constant and non-constant curvature case.

\subsection{LRS Bianchi Type $I$ Spacetime}

The line element of LRS Bianchi type $I$ spacetime is
given by
\begin{equation}\label{6}
ds^{2}=dt^2-A^2(t)dx^2-B^2(t)[dy^2+dz^2],
\end{equation}
where $A$ and $B$ are cosmic scale factors. The corresponding
Ricci scalar turns out to be
\begin{equation}\label{7}
R=-2[\frac{\ddot{A}}{A}+\frac{2\ddot{B}}{B}+
\frac{2\dot{A}\dot{B}}{AB}+\frac{\dot{B}^2}{B^2}],
\end{equation}
where dot denotes derivative with respect to $t$.
The average scale factor $a$ and the volume scale factor $V$ are
defined as
\begin{equation}\label{102}
a=\sqrt[3]{AB^2},\quad V=a^3=AB^2.
\end{equation}
The average Hubble parameter $H$ is given in the form
\begin{equation}\label{103}
H=\frac{1}{3}(\frac{\dot{A}}{A}+\frac{2\dot{B}}{B}).
\end{equation}
The expansion scalar $\theta$ and shear scalar $\sigma$ are
defined as follows
\begin{eqnarray}\label{104}
\theta&=&u^\mu_{;\mu}=\frac{\dot{A}}{A}+2\frac{\dot{B}}{B},\\
\label{15} \sigma^2&=&\frac{1}{2}\sigma_{\mu\nu}\sigma^{\mu\nu}
=\frac{1}{3}[\frac{\dot{A}}{A}-\frac{\dot{B}}{B}]^2,
\end{eqnarray}
where
\begin{equation}\label{106}
\sigma_{\mu\nu}=\frac{1}{2}(u_{\mu;\alpha}h^\alpha_\nu+u_{\nu;\alpha}h^\alpha_\mu)
-\frac{1}{3}\theta h_{\mu\nu},
\end{equation}
$h_{\mu\nu}=g_{\mu\nu}-u_{\mu}u_{\nu}$ is the projection tensor.
Now we explore the solutions of the field equations for two classes of $f(R, T)$ models.\\

\subsection{$f(R,T)=R+2f(T)$}

For the model $f(R,T)=R+2f(T)$, the field equations become
\begin{equation}\label{11}
R_{\mu\nu}-\frac{1}{2}Rg_{\mu\nu}=\kappa
T_{\mu\nu}+2f_T(T)T_{\mu\nu}+\bigg[f(T)+2pf_T(T)\bigg]g_{\mu\nu}.
\end{equation}
Here we find the most basic possible solution of this
theory due to the complicated nature of field equations.
For the sake of simplicity, we use
natural system of units $(G=c=1)$ and $f(T)=\lambda T$, where
$\lambda$ is an arbitrary constant. In this case, the gravitational field equations take the form similar to GR
\begin{equation}\label{12}
R_{\mu\nu}-\frac{1}{2}Rg_{\mu\nu}-\lambda (T+2p)g_{\mu\nu}=(8\pi+2\lambda)T_{\mu\nu},
\end{equation}
where the term $\lambda(T+2p)$ may play the role of cosmological constant $\Lambda$ of the GR  field equations.
It would be worthwhile to mention here that the dependence of cosmological constant $\Lambda$ on the trace of energy momentum tensor $T$ has already been proposed by Poplawski \cite{Poplawski} and the cosmological constant in the gravitational Lagrangian is a function of $T$. Consequently the model was named as ``$\Lambda(T)$ gravity". It has been proved
that recent astrophysical data favor a variable cosmological constant which is consistent with $\Lambda(T)$ gravity.
$\Lambda(T)$ gravity has been shown to be more general than the Palatini $f(R)$ gravity \cite{Poplawski1}.
Now using Eq. (\ref{12}), we obtain a set of differential equations for LRS Bianchi type $I$ spacetime,
\begin{eqnarray} \label{13}
\frac{2\dot{A}\dot{B}}{AB}+\frac{\dot{B}^2}{B^2}&=&(8\pi+3\lambda)\rho-\lambda p,\\\label{14}
-\frac{2\ddot{B}}{B}-\frac{\dot{B}^2}{B^2}&=&(8\pi+3\lambda)p-\lambda\rho,\\\label{15}
-\frac{\ddot{A}}{A}-\frac{\ddot{B}}{B}-\frac{\dot{A}\dot{B}}{AB}&=&(8\pi+3\lambda)p-\lambda\rho.
\end{eqnarray}
Thus we have three differential equations with four unknowns namely $A,~B$, $p$ and $\rho$.
Adding Eqs. (\ref{13}) and (\ref{14}) gives
\begin{equation}\label{16}
\frac{2\dot{A}\dot{B}}{AB}-\frac{2\ddot{B}}{B}=(8\pi+2\lambda)(\rho+p).
\end{equation}
Similarly, addition of Eqs. (\ref{13}) and (\ref{15}) yields
\begin{equation}\label{17}
\frac{\dot{A}\dot{B}}{AB}-\frac{\ddot{A}}{A}-\frac{\ddot{B}}{B}
+\frac{\dot{B}^2}{B^2}=(8\pi+2\lambda)(\rho+p).
\end{equation}
Subtracting Eqs.(\ref{16}) and (\ref{17}), it follows that
\begin{equation}\label{18}
\frac{\dot{A}\dot{B}}{AB}+\frac{\ddot{A}}{A}-\frac{\ddot{B}}{B}
-\frac{\dot{B}^2}{B^2}=0.
\end{equation}
Now we are left with only one differential equation and two unknowns.
Therefore, we need an additional constraints to Eq.(\ref{18}).
Here we use a physical condition that expansion scalar $\theta$ is
proportional to shear scalar $\sigma$ which provides
\begin{equation}\label{19}
A=B^n,
\end{equation}
where $n$ is an arbitrary real number and we consider $n\neq 0,1$ for non-trivial solutions.
The physical reason for this assumption is justified as the observations of the velocity red-shift relation for extragalactic sources suggest that Hubble expansion of the universe may achieve isotropy when $\frac{\sigma}{\theta}$ is constant \cite{Phyreason1}. Collins \cite{Phyreason2} provided the physical significance of this condition for perfect fluid with barotropic EoS. In literature \cite{21}, \cite{Phyreason3}-\cite{Phyreason7}, many authors have proposed this condition
to find the exact solutions of field equations.

Thus using Eq.(\ref{19}), Eq.(\ref{18}) take the form
\begin{equation}\label{20}
\frac{\ddot{B}}{B}+(n+1)\frac{\dot{B}^2}{B^2}
=0,
\end{equation}
which yields a solution
\begin{equation}\label{21}
B=c_1[(n+2)t+c_2]^{\frac{1}{n+2}},
\end{equation}
where $c_1$ and $c_2$ are constants of integration.
Thus, the solution metric takes the form
\begin{equation}\label{22}
ds^{2}=dt^2-{c_1}^{2n}[(n+2)t+c_2)]^{\frac{2n}{n+2}}dx^2-{c_1}^2[(n+2)t+c_2)]^{\frac{2}{n+2}}[dy^2+dz^2].
\end{equation}
The volume scale factor turn out to be
\begin{equation}\label{22a}
V=a^3={c_1}^{n+2}[(n+2)t+c_2].
\end{equation}
The expansion scalar and the shear scalar become
\begin{equation}\label{22b}
\theta=\frac{n+2}{(n+2)t+c_2},\quad \sigma^2=\frac{1}{3}\bigg[\frac{n-1}{(n+2)t+c_2}\bigg]^2.
\end{equation}
It is mentioned here that the isotropy condition, i.e.,
$\frac{\sigma^2}{\theta}\rightarrow 0$ as $t\rightarrow \infty$,
is satisfied in this case. It can be observed from Eqs. (\ref{22a}) and (\ref{22b}) that the spatial volume is zero at $t=0$ while the expansion scalar is infinite, which suggests that the universe starts evolving with zero volume at $t=0$, i.e. big bang scenario.
It is further observed that the average scale factor is zero at the initial epoch $t=0$ and hence the model has a
point type singularity \cite{MacCallum}.
The energy density and pressure of the universe take the form
\begin{equation}
\rho=p=\frac{(1+2n)}{2(\lambda+4\pi)[(n+2)t+c_2]^2},
\end{equation}
which suggest that equation of state parameter $\omega=1$ corresponding to stiff fluid universe.
The average Hubble parameter turn out to be
\begin{equation}
H=\frac{n+2}{3[(n+2)t+c_2]}.
\end{equation}
Therefore
\begin{equation}\label{202c}
\frac{H}{H_0}=\frac{(n+2)t_0+c_2}{(n+2)t+c_2},
\end{equation}
where $H_0$ is the present value of Hubble's parameter. The redshift for a distant source is directly related to the scale factor of the universe at the time when the photons were emitted from the source. The scale factor $a$ and redshift $z$ are related through the equation
\begin{equation}
a=\frac{a_0}{1+z},
\end{equation}
where $a_0$ is the present value of the scale factor. Thus we obtain
\begin{equation}\label{22d}
\frac{a_0}{a}=1+z=\bigg[\frac{(n+2)t_0+c_2}{(n+2)t+c_2}\bigg]^{\frac{1}{3}}.
\end{equation}
Using Eqs. (\ref{202c}) and (\ref{22d}), we obtain the value of Hubble's parameter in terms of redshift parameter
\begin{equation}
H=H_0(1+z)^3.
\end{equation}
According to the Hubble Law, the distance of a given galaxy is proportional to
the recessional velocity as measured by the doppler red shift. Thus, the
value of Hubble's parameter in terms of redshift has much importance in astrophysical contexts.

The deceleration parameter $q$ in cosmology is the
measure of the cosmic acceleration of the universe expansion and
is defined as
\begin{equation}
q=-\frac{\ddot{a}a}{\dot{a}^2}.
\end{equation}
It is mentioned here that the behavior of the
universe models depend upon the sign of $q$. The positive
deceleration parameter provides a decelerating model while the
negative value corresponds to inflation. For this solution, the value of deceleration parameter turns out to be
$q=2$ which suggests a decelerating model of universe.

Universe models closed to $\Lambda$CDM can be described using the cosmic jerk parameter $j$, a dimensionless
third derivative of the scale factor with respect to the cosmic time \cite{Visser}. The value of jerk parameter is constant for flat $\Lambda$CDM model. The jerk parameter is defined as
\begin{equation}
j=\frac{1}{H^3}\frac{\dot{\ddot{a}}}{a}.
\end{equation}
The expression for jerk parameter in terms of deceleration parameter turns out to be
\begin{equation}
j=q+2q^2-\frac{\dot{q}}{H}.
\end{equation}
Thus we obtain $j=10$ in the case of our solution.
It would be worthwhile to mention here that this solution gives
$R=0$ for $n=-\frac{1}{2}$. In this case, the solution metric takes the form
\begin{equation}\label{22}
ds^{2}=dt^2-[c_1(\frac{3}{2}t+c_2)]^{-\frac{2}{3}}dx^2-[c_1(\frac{3}{2}t+c_2)]^{\frac{4}{3}}[dy^2+dz^2].
\end{equation}
Without loss of generality, we take $c_2=0$ and re-define the parameters, i.e., $\sqrt[3]{{\frac{2x^3}{3{c_1}^{3/2}}}}\longrightarrow
\tilde{x},~\sqrt[3]{{\frac{9{c_1}^3y^3}{4}}}\longrightarrow \tilde{y}$ and
$\sqrt[3]{{\frac{9{c_1}^3z^3}{4}}}\longrightarrow \tilde{z}$, the above metric takes the
form
\begin{equation}\label{42a}
ds^{2}=dt^{2}-t^{-\frac{2}{3}}d\tilde{x}^{2}-t^{\frac{4}{3}}(d\tilde{y}^{2}+d\tilde{z}^{2}),
\end{equation}
which is exactly the same as the well-known Kasner's metric \cite{30}.\\\\
Now we discuss the possibility of solutions for a non-linear forms of $f(T)$.
We assume $f(T)=\lambda T^2$ so that Eq. (\ref{11}) takes the form
\begin{equation}\label{1002}
R_{\mu\nu}-\frac{1}{2}Rg_{\mu\nu}=(8\pi+4\lambda T)T_{\mu\nu}+\lambda T(T+4p)g_{\mu\nu}.
\end{equation}
Now using Eq. (\ref{1002}), we obtain a set of independent differential equations for LRS Bianchi type $I$ spacetime,
\begin{eqnarray} \label{1003}
\frac{2\dot{A}\dot{B}}{AB}+\frac{\dot{B}^2}{B^2}&=&8\pi\rho+5\lambda{\rho}^2-14\lambda\rho p-3\lambda p^2,\\\label{1004}
-\frac{2\ddot{B}}{B}-\frac{\dot{B}^2}{B^2}&=&8\pi p-9\lambda p^2+6\lambda p\rho-\lambda {\rho}^2,\\\label{1005}
-\frac{\ddot{A}}{A}-\frac{\ddot{B}}{B}-\frac{\dot{A}\dot{B}}{AB}&=&8\pi p-9\lambda p^2+6\lambda p\rho-\lambda {\rho}^2.
\end{eqnarray}
These equations yield the same solution metric as given by Eq. (\ref{22}).
However, in this case the energy density can be obtained by solving the equation
\begin{eqnarray}
\rho^2+\frac{3-\omega}{2\lambda(1-3\omega)}\rho-\frac{1+2n}{8\lambda(1-3\omega)[(n+2)t+c_2]^2}=0.
\end{eqnarray}
The quadratic equation is due to non-linear form of $f(T)$.

\subsection{$f(R,T)=f_1(R)+f_2(T)$}

 Now we explore the solutions with more general class. Here the field equations for the model $f(R,T)=f_1(R)+f_2(T)$ become
\begin{equation}\label{2129}
{f_1}_R(R)R_{\mu\nu}-\frac{1}{2}f_1(R)g_{\mu\nu}-(\nabla_{\mu}
\nabla_{\nu}-g_{\mu\nu}\Box){f_1}_R(R)=[\kappa+{f_2}_T(T)]T_{\mu\nu}+[p{f_2}_T(T)+\frac{1}{2}f_{2}(T)]g_{\mu\nu}.
\end{equation}
Contracting the field equations (\ref{2129}), we obtain
\begin{equation}\label{2229}
R{f_1}_R(R)-2f_1(R)+3\Box{f_1}_R(R)=\kappa
T+2f_{2}(T)+[T+4p]{f_2}_T(T).
\end{equation}
Using this, we can write
\begin{equation}
f_1(R)=\frac{3\Box{f_1}_R(R)+R{f_1}_R(R)-\kappa
T-2f_{2}(T)-[T+4p]{f_2}_T(T)}{2}.
\end{equation}
Inserting this in Eq.(\ref{2129}), we get
\begin{eqnarray}\nonumber
\frac{{f_1}_R(R)R_{\mu\nu}-\nabla_{\mu}
\nabla_{\nu}{f_1}_R(R)-(\kappa+{f_2}_T(T))T_{\mu\nu}}{g_{\mu\nu}}=\\\label{23}\frac{R{f_1}_R(R)-\Box{f_1}_R(R)-\kappa T-T{f_2}_T(T)}{4}.
\end{eqnarray}
Since the metric (\ref{6}) depends only on $t$, one can view
Eq.(\ref{23}) as the set of differential equations for ${f_1}_R(t),~{f_2}_T(t)$, $A$, $B$, $\rho$ and $p$. It follows from Eq.(\ref{23}) that the combination
\begin{equation}\label{2429}
A_{\mu}\equiv\frac{{f_1}_R(R)R_{\mu\mu}-\nabla_{\mu}\nabla_{\mu}
{f_1}_R(R)-(\kappa+{f_2}_T(T))T_{\mu\nu}}{g_{\mu\mu}},
\end{equation}
is independent of the index $\mu$ and hence $A_{\mu}-A_{\nu}=0$
for all $\mu$ and $\nu$. Thus $A_{0}-A_{1}=0$ yields
\begin{equation}\label{3529}
-\frac{2\ddot{B}}{B}+\frac{2\dot{A}\dot{B}}{AB}
+\frac{\dot{A}{\dot{{f_1}_R}(R)}}{A{f_1}_R(R)}-\frac{{\ddot{{f_1}_R}(R)}}{{f_1}_R(R)}-\bigg[\frac{\kappa+{f_2}_T(T)}{{f_1}_R(R)}\bigg](\rho+p)=0.
\end{equation}
Also, $A_{0}-A_{2}=0$ provides
\begin{equation}\label{3629}
-\frac{\ddot{A}}{A}-\frac{\ddot{B}}{B}
+\frac{\dot{A}\dot{B}}{AB}+\frac{\dot{B}^2}{B^2}
+\frac{\dot{B}{\dot{{f_1}_R}(R)}}{B{f_1}_R(R)}-\frac{{\ddot{{f_1}_R}(R)}}{{f_1}_R(R)}-\bigg[\frac{\kappa+{f_2}_T(T)}{{f_1}_R(R)}\bigg](\rho+p)
\end{equation}
Now we have two differential equations with six unknowns namely $A,~B$, ${f_1}(R)$, ${f_2}(T)$, $p$ and $\rho$.
Here we also use $A=B^n$ so that Eqs. (\ref{3529}) and (\ref{3629}) take the form
\begin{eqnarray} \label{3729}
-\frac{2\ddot{B}}{B}+2n\frac{\dot{B}^2}{B^2}
+n\frac{\dot{B}\dot{{f_1}_R}(R)}{B{f_1}_R(R)}-\frac{\ddot{{f_1}_R}(R)}{{f_1}_R(R)}-
\bigg[\frac{\kappa+{f_2}_T(T)}{{f_1}_R(R)}\bigg](\rho+p)=0,\\\label{3829}
(n+1)\frac{\ddot{B}}{B}+(n^2-2n-1)\frac{\dot{B}^2}{B^2}-
\frac{\dot{B}\dot{{f_1}_R}(R)}{B{f_1}_R(R)}+\frac{\ddot{{f_1}_R}(R)}{{f_1}_R(R)}+\bigg[\frac{\kappa+{f_2}_T(T)}{{f_1}_R(R)}\bigg](\rho+p)=0.
\end{eqnarray}
\textbf{Case I: Exponential Law Solutions}\\\\
It has been proved that dark matter and dark energy phases
can be achieved by finding the exact solutions using a power
law $f(R)$ model \cite{02}. So it would be interesting to assume
$f_1(R)$ in power law form to solve the field equations. We follow the approach of Nojiri and
Odintsov \cite{Unified cosmic history in modified gravity from
F(R) theory to Lorentz non-invariant models} and take the
assumption ${f_1}_R(R)\propto f_0R^m$, where $f_0$ is an arbitrary
constant.  So in this case the addition of Eqs.(\ref{3729},\ref{3829}) yields
\begin{eqnarray}\label{24290}
&&(n^3+2n^2+2n+2)\dot{B}^4+(2n^2+4n+3)B\dot{B}^2\ddot{B}+(n+2)B^2{\ddot{B}}^2-\nonumber\\&&
2m(n^2+n+1)\dot{B}^4+m(2n^2+n)B\dot{B}^2\ddot{B}+m(n+2)B^2\ddot{B}\dddot{B}=0.
\end{eqnarray}
It is interesting to notice that this equation admits an exponential solution of the form
\begin{equation}\label{25290}
B(t)=e^{c_3t+c_4},
\end{equation}
where $c_3$ and $c_4$ are arbitrary constants. The exponential solution is satisfied with the constraint equation
\begin{equation}\label{26290}
n^3+4n^2+7n+6=0.
\end{equation}
The solutions of this equation turn out to be
\begin{equation}
n=-2,\quad -1\pm i\sqrt{2}.
\end{equation}
It is mentioned here that the real value of $n$ gives a constant Ricci scalar
while we obtain a non-constant Ricci scalar for the complex values of $n$. We discard the imaginary case and consider the real value of $n$ to get a physical solution
\begin{equation}\label{27290}
ds^{2}=dt^2-e^{-4(c_3t+c_4)}dx^2-e^{2(c_3t+c_4)}(dy^2+dz^2).
\end{equation}
The average Hubble parameter turn out to be zero here. All other dynamical quantities
like volume scale factor of universe, expansion scalar $\theta$ and shear scalar
$\sigma$ are constant for this solution. Energy density
and pressure of the universe are related by the equation
\begin{eqnarray}
\rho+p=\frac{(-6)^{m+1}f_0{c_3}^2}{\kappa+{f_2}_T(T)}.
\end{eqnarray}
Many expressions for pressure and energy density can be evaluated for different choices of ${f_2}(T)$.
For example when ${f_2}(T)=\lambda T^2$ and using Eq.(\ref{eos}), we obtain
\begin{eqnarray}
\rho^2+\frac{\kappa}{2\lambda(1-3\omega)}\rho-\frac{(-6)^{m+1}f_0{c_3}^2}{2\lambda(1+\omega)(1-3\omega)}=0
\end{eqnarray}
which is quadratic in $\rho$ and one can work out its roots to get energy density.\\\\
\textbf{Case II: Power Law Solutions}\\\\
Here we assume that the solution is in power law form, i.e.
$B(t)=(c_5t+c_6)^k$, where $c_5,~c_6$ and $k$ are arbitrary real constants with $k\neq 0$.
Using Eq.(\ref{24290}), we obtain a constraint equation
\begin{equation}\label{28290}
(n^3+4n^2+7n+6)k^2-[2m(n^2+2n+3)+2n^2+6n+7]k+(1+2m)(n+2)=0.
\end{equation}
This equation is important because it will be used to reconstruct different forms
of $f_1(R)$ models with suitable solutions of field equations.
For example, here we investigate the solution for $n=-1$. In this case, Eq.(\ref{28290}) gives
\begin{equation}\label{29290}
k=2m+1,\quad m\neq 0,~1.
\end{equation}
So the solution metric takes the form
\begin{equation}\label{30290}
ds^{2}=dt^2-(c_5t+c_6)^{-2(2m+1)}dx^2-(c_5t+c_6)^{2(2m+1)}(dy^2+dz^2).
\end{equation}
The volume scale factor and average Hubble parameter become here
\begin{equation}\label{65290}
V=a^3=(c_5t+c_6)^{2(2m+1)},\quad H=\frac{c_5(2m+1)}{3(c_5t+c_6)}.
\end{equation}
The expansion scalar and shear scalar turn out to be
\begin{equation}\label{66290}
\theta=\frac{c_5(2m+1)}{c_5t+c_6},\quad \sigma^2=\frac{4}{3}[\frac{c_5(2m+1)}{c_5t+c_6}]^2.
\end{equation}
The isotropy condition $\frac{\sigma^2}{\theta}\rightarrow 0$ as $t\rightarrow \infty$,
is also satisfied in this case.
Using Eq.(\ref{65290}), we get
\begin{equation}\label{22c290}
\frac{H}{H_0}=\frac{c_5t_0+c_6}{c_5t+c_6},~~~~\frac{a_0}{a}=1+z=\bigg[\frac{c_5t_0+c_6}{c_5t+c_6}\bigg]^{\frac{2(2m+1)}{3}}.
\end{equation}
Thus the value of Hubble's parameter in terms of redshift parameter turns out to be
\begin{equation}\label{refree1}
H=H_0(1+z)^{\frac{3}{2(2m+1)}}.
\end{equation}
The deceleration parameter in this case becomes
\begin{equation}
q=\frac{1-4m}{2(2m+1)}
\end{equation}
while the jerk parameter is given by
\begin{equation}\label{refree2}
j=\frac{8(4m^2-5m+1)}{(2m+1)^2}.
\end{equation}
By observing Eqs.(\ref{refree1}-\ref{refree2}), it is clear that singularity occurs at $m=-\frac{1}{2}$. 
Further, the Ricci scalar turns out to be
\begin{equation}\label{69290}
R=-\frac{2{c_5}^2(1+2m)(1+4m)}{(c_5t+c_6)^2}.
\end{equation}
For $m=-\frac{1}{2}$ or $m=-\frac{1}{4}$, the Ricci scalar turns out to be constant, i.e. $R=0$.
$m=-\frac{1}{2}$ corresponds to Minkwoski spacetime while
for $m=-\frac{1}{4}$ the metric takes the form
\begin{equation}\label{30290}
ds^{2}=dt^2-(c_5t+c_6)^{-1}dx^2-(c_5t+c_6)(dy^2+dz^2).
\end{equation}
This solution gives a point singularity at $t=-\frac{c_6}{c_5}$.
The Ricci scalar remains non-constant for $m\neq -\frac{1}{2},-\frac{1}{4}$.
For a special case when $m=-1$, $f_1(R)$ takes the logarithmic form
\begin{equation}\label{68290}
f_1(R)\propto f_0\ln|R|+c_7,
\end{equation}
where $c_7$ is an integration constant. Here energy density and pressure of the universe are related by the equation
\begin{eqnarray}
\rho+p=-\frac{f_0}{\kappa+{f_2}_T(T)}.
\end{eqnarray}
Here we can also calculate many expressions for energy density depending upon the value of ${f_2}(T)$.

\section{Concluding Remarks}

This paper is devoted to study Bianchi type
cosmology in $f(R,T)$ gravity. We explore the exact
solutions of field equations for LRS Bianchi type $I$
spacetime. Since the field equations are highly nonlinear and complicated,
we use the assumption that the expansion scalar $\theta$ is
proportional to the shear scalar $\sigma$ to solve them. It gives $A=B^n$, where
$A,~B$ are the metric coefficients and $n$ is an arbitrary
constant. Mainly we have explored three solutions of modified  field equations using different assumptions.

The first solution is obtained for the model $f(R,T)=R+2f(T)$.
The isotropy condition, i.e. $\frac{\sigma^2}{\theta}\rightarrow 0$ as $t\rightarrow \infty$,
is satisfied for the solution. The spatial volume is zero at $t=0$ and the expansion scalar is infinite, which suggests that the universe starts evolving with zero volume at $t=0$, i.e. big bang scenario.
The average scale factor turns out to be zero at the initial epoch $t=0$ and hence the model has a
point type singularity \cite{MacCallum}. The expressions for energy density and pressure suggest that EoS parameter $\omega=1$ which corresponds to stiff fluid universe. The deceleration parameter $q$ turns out to be
$q=2$ which suggests a decelerating model of universe. We have also calculated jerk parameter $j=10$ in the case of our solution.
It is worth mentioning here that this solution gives
$R=0$ for $n=-\frac{1}{2}$ corresponds to well-known Kasner's solution already available in GR \cite{30}.

We have also explored the more general solutions of field equation by considering the model
$f(R,T)=f_1(R)+f_2(T)$. Moreover, we have not used any conventional assumption like constant deceleration parameter or variation law of Hubble's parameter to investigate the solutions in this case. In particular, exponential law and power law solutions have been investigated for this model.
Foe exponential law case, the average Hubble parameter turn out to be zero and all other dynamical quantities
like volume scale factor of universe, expansion scalar $\theta$ and shear scalar
$\sigma$ are constant. It would be worthwhile to mention here that when $f_2(T)=0$, this class corresponds to
$f(R)$ gravity model. For $f_2(T)\neq 0$, different expressions for energy density can be generated with different choices of $f_2(T)$ models.
Power law solution provides a non-constant scalar curvature and
thus many important $f(R)$ models can be reconstructed. As a special case, we have developed an important logarithmic $f(R)$ model.\\\\\\\\
\textbf{Acknowledgement}\\\\ The author is thankful to National University
of Computer and Emerging Sciences (NUCES) for
funding support through research reward programme.
The author is also grateful to the anonymous reviewer
for valuable comments and suggestions to improve the paper.

\end{document}